\documentclass[twocolumn,showpacs,amsmath,amssymb,prb]{revtex4}

\usepackage{graphicx}
\usepackage{dcolumn}
\usepackage{bm}

\def\be{\begin{eqnarray}}
\def\ee{\end{eqnarray}}
\def\ba{\begin{array}}
\def\ea{\end{array}}

\begin{document}

\preprint{1}

\title{Thomas-Fermi-Poisson theory of screening for laterally confined and
  unconfined two-dimensional electron systems in strong magnetic fields}

\author{A. Siddiki}%
\author{Rolf R. Gerhardts}
\affiliation{Max-Planck-Institut f\"ur Festk\"orperforschung, 
Heisenbergstrasse 1, D-70569 Stuttgart, Germany}%

\date{\today}

\begin{abstract}
 We examine within the self-consistent Thomas-Fermi-Poisson approach
 the low-temperature screening properties of a two-dimensional 
electron gas (2DEG)  subjected to strong perpendicular 
magnetic fields. Numerical results for the unconfined 2DEG are compared with
 those for a simplified Hall bar geometry realized by two different
 confinement models. It is shown that in the strongly non-linear screening
 limit of zero temperature the total variation of the screened potential is
 related by simple analytical expressions to the amplitude of an applied
 harmonic modulation potential and to the strength of the magnetic field.
\end{abstract}

\maketitle

\section{\label{sec:1} Introduction}

A two-dimensional electron gas (2DEG) in a strong perpendicular magnetic field
has unusual low-temperature screening properties,
\cite{Wulf88:4218,Efros88:1019} since the highly degenerate Landau-quantized
energy levels lead to a strong variation of the thermodynamic density of
states (TDOS) with varying strength of the magnetic field, i.e., with varying
filling factor $\nu$ of the Landau levels (LLs). If a LL is close to half
filled, the TDOS is very high (inversely proportional to the temperature
$T$), and static potential fluctuations are nearly perfectly screened. We will
consider only spin-degenerate 2DEGs,  
so that this happens if the value of $\nu$ is close to an odd integer, 
while at even-integer  $\nu$ the Fermi energy lies in the
gap between two adjacent LLs and a spatial redistribution of
electrons and, therefore, a screening of (weak) potential fluctuations is
impossible. 
 In an inhomogeneous 2DEG with sufficiently strong long-range density
 fluctuations,  screening effects lead to quasi metallic (so called
``compressible'') regions with high TDOS, in which screening is nearly perfect
and a LL is ``pinned'' to the Fermi energy, and to insulator-like
``incompressible'' regions, which separate adjacent compressible regions.
 In the incompressible regions the Fermi energy falls into the gap between two
 LLs and the electron density $n_{\rm el}({\bf r})$ is
 constant (even-integer filling factor), while in the compressible regions
$n_{\rm el}({\bf r})$ adjusts itself so that the self-consistent electrostatic 
potential energy $V({\bf r})$ of an electron differs from the Fermi energy
(more precisely  the  electrochemical potential $\mu^{\star}$) by a
Landau energy $\hbar \omega_c 
(n+1/2)$, where $\omega_c=eB/m$ is the cyclotron frequency in the magnetic
field $B$. As a consequence, $V({\bf r})$ becomes nearly constant within a
compressible region and differs by integer multiples of  $\hbar \omega_c$
between different compressible regions.
Landau level pinning and the interplay of
compressible and incompressible regions  lead to strongly nonlinear
screening effects. 
This screening scenario has been established some time ago
\cite{Wulf88:4218,Efros88:1019}
and was applied, e.g., to calculate, at zero temperature, the electronic DOS
\cite{Efros93:2233} and transport \cite{Burnett93:14365,Cooper93:4530} through
2DEGs  in smooth periodic and random potentials.
The explanation \cite{Tsemekhman97:R10201,Guven02:155316} of several
experimental results, e.g., on quantum Hall devices 
under high currents close to the breakdown of the quantized Hall effect,
\cite{Kaya99:62,Kaya00:128}
rely also on these ideas. A systematic
investigation of these interesting nonlinear screening effects is, however,
apparently not available in the literature.

Models for half-space and Hall-bar geometries with planar charge
distributions have been proposed
that allow closed solutions of Poisson's equation (i.e., the
calculation of the potential for given electron density),  and
estimates of position and widths of the incompressible strips have been given.
 \cite{Chklovskii92:4026,Chklovskii93:12605} By adding  the
non-linear Thomas-Fermi approximation for the calculation of the electron
density from the potential,
 that work was  extended to a self-consistent approach,
which allows us to calculate both electron density and electrostatic potential
 for arbitrary temperature. \cite{Lier94:7757,Oh97:13519} This approach, which
 we will employ in the following, shows
 that the existence and the width of 
incompressible strips depends sensitively on temperature, and allows to
calculate their position and width for given background charges without
additional assumptions.

The purpose of the present work is a systematic investigation of the nonlinear
low-temperature screening of harmonic electrostatic potential modulations in
laterally confined and unconfined 2DEGs   
subjected to a quantizing perpendicular magnetic field. We will demonstrate
that in general edge effects do not qualitatively change the screening
properties of the 2DEG, even if the sample width is not much larger than the
period of the imposed potential modulation. There are, however, peculiar
differences between confined and unconfined 2DEGs in situations, in which the
latter have no states near the Fermi energy. To understand this in detail, we
first discuss the screening of a potential modulation imposed on a homogeneous
2DEG (Sec.II) and then consider, for two different boundary models, edge
effects on screening in Hall bar geometries (Sec.III).

 We will assume the
 2DEG to be located in the plane $z=0$ with a (surface)  number density
 $n_{\rm el}(x)$ and consider only situations with translation invariance in the
 $y$ direction. The (Hartree)
 contribution $V_H(x)$  to the potential energy of an electron caused by the
total charge density of the 2DEG can be written as \cite{Oh97:13519}
\be  \label{hartree}
V_H(x)= \frac{2e^2}{\bar{\kappa}} \int_{x_l}^{x_r} \!dx' K(x,x')\,n_{\rm
  el}(x'), 
\ee
where $-e$ is the electron charge, $\bar{\kappa}$ an average background
dielectric constant, \cite{Oh97:13519} and the kernel $K(x,x')$ describes the
solution of Poisson's equation with appropriate boundary conditions at $x_l$
and $x_r$.
The electron density in turn is calculated in the 
 Thomas-Fermi  approximation (TFA) \cite{Oh97:13519}
\be \label{thomas-fermi}
 n_{\rm el}(x)=\int dE\,D(E)f\big( [E+V(x)-\mu^{\star}]/k_{B}T \big),
\ee
with $D(E)$ the relevant (single-particle) density of states (DOS),
$f(\epsilon)=[1+e^{\epsilon}]^{-1}$ the Fermi function, $\mu^{\star}$ the
electrochemical potential, and with $V(x)=V_{\rm ext}(x)+V_H(x)$ the total
potential energy of an electron, which differs from $V_H(x)$ by the
contribution due to external charges, e.g., a homogeneous positively charged
background and a charge distribution creating a periodic modulation potential.
The local (but nonlinear) TFA is much simpler than the corresponding quantum
mechanical calculation and expected to yield essentially the same results if
$V(x)$ varies slowly in space, i.e., on a length scale much larger than
typical quantum lengths such as the extent of wave functions or the Fermi
wavelength.  

\section{\label{sec:2} Homogeneous 2DEG} 

We start with a homogeneous 2DEG described by the DOS $D_0(E)=D_0 \theta(E)$,
with $D_0=m/(\pi \hbar^2)$, for $B=0$, and by the Landau DOS
\be \label{landau-dos}
D_B(E)=\frac{1}{\pi 
l_{m}^{2}}\sum_{n=0}^{\infty}{\delta (E-E_n)},
\quad E_n=\hbar\omega_{c}(n+1/2)
 \ee
for finite $B$. The
effective mass of an electron is denoted by $m$, the  magnetic length by
$l_{m}=\sqrt{\hbar /(m \omega_c)}$. In both cases we assume spin degeneracy
and neglect 
collision broadening effects. The constant electron density $\bar{n}_{\rm el}$
is given by Eq.~(\ref{thomas-fermi}) with $\bar{n}_{\rm el}=n_{\rm el}(x)$,
$V(x)\equiv 0$ and $\mu^{\star}=\mu$ the chemical potential.
For these simple models the energy integral in Eq.~(\ref{thomas-fermi}) is
readily carried out. We tacitly assume that the electron charges are
neutralized by a homogeneous background of positive charges.

\subsection{Kernel for periodic modulation \label{sec:2-kernel}}

We now add a periodic external modulation described by a potential energy
$V_{\rm ext}(x)=V_{\rm ext}(x+a)$. The 2DEG will respond with a density
modulation and a Hartree potential of the same period $a$. To exploit the
periodicity, we expand density and potentials into Fourier series according to
\be \label{fourier}
V(x)=\sum_q V^q \, e^{iqx}, \quad  V^q=\int_{-a/2}^{a/2}\! \frac{dx}{a}\,
e^{-iqx}V(x), 
 \ee
with $q=2\pi n/a$ and integer $n$. To maintain charge neutrality, we require
$n_{\rm el}^0=\bar{n}_{\rm el}$ in any case. With the boundary conditions
$V_H(x,z)\rightarrow 0$ for $|z|\rightarrow \infty$, Poisson's equation yields
(see, e.g., Ref.~\onlinecite{Ando82:437})
\be \label{harmonic}
 V_H^q(z)= (2\pi e^2/\bar{\kappa} |q|)e^{-|qz|}\,n_{\rm el}^q 
\ee
as response  to the density fluctuation $n_{\rm el}^q$.
Summing over harmonics (for $q\neq 0$),\cite{Morse-Feshbach53:1240}  we obtain
 $V_H(x,z=0)$ from  Eq.~(\ref{hartree}) with $-x_l=x_r=a/2$ and the
kernel
\be \label{kernel-hom}
K(x,x')=- \ln \left| 2\sin \, \frac{\pi}{a} (x-x')
\right| \,.
\ee
%
 
\subsection{Breakdown of linear screening \label{sec:2-breakdown}}
\subsubsection{Zero magnetic field}

\begin{figure}
{\centering \includegraphics[width=0.9\linewidth]{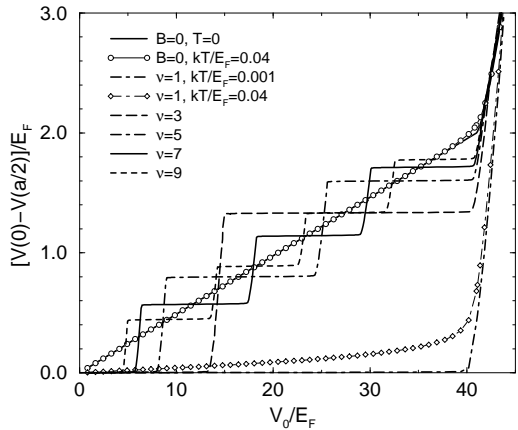}}
\caption{ \label{fig:odd}
Variance of the screened potential versus amplitude of the  harmonic
potential modulation imposed on a spin-degenerate homogeneous 2DEG with a half
filled Landau level, for several odd-integer values of the filling factor.
Default temperature $k_BT/E_F=0.001$, $\epsilon(q)=41$. }
\end{figure}

In the limit $B=0$, $T\rightarrow 0$ and with $E_F=\mu^{\star}(B=0,T=0)$,
Eq.~(\ref{thomas-fermi}) reduces to 
\be \label{linear-resp}
n_{\rm el}(x)=D_0\,\big(E_F-V(x)\big) \,
\theta\big(E_F-V(x)\big) , 
\ee
which is a linear relation between $V(x)$ and $n_{\rm el}(x)$ for
$V(x)< E_F$. With Eq.~(\ref{harmonic}) we find for a harmonic potential
modulation $V_{\rm ext}(x)=V^q_{\rm ext}\cos qx$ a harmonic density
modulation $\delta n_{\rm el}(x)=n^q_{\rm el}\cos qx$ and 
the self-consis\-tent (``screened'') potential $V(x)=V^q \cos qx$ with 
\be \label{linear-eps}
V^q =V^q_{\rm ext} /\epsilon (q), \quad \epsilon (q)=1+Q_0/|q|.
\ee
The dielectric function $\epsilon (q)$ can be expressed in terms of the
effective Bohr radius $a_{\rm B}^{\star}=\bar{\kappa} \hbar^2 /(m e^2)$
(for GaAs $a_{\rm B}^{\star} =9.8\,$nm),  since
 $Q_0=2 \pi e^2 D_0/\bar{\kappa}=2/a_{\rm
   B}^{\star}$. \cite{Stern67:546,Wulf88:4218} With $q=2\pi/a$, the screening
 strength is thus determined by the dimensionless parameter 
\be \label{alfascr}
\alpha=\pi a_{\rm B}^{\star}/a .
\ee
We will assume $\alpha \ll 1$, i.e., $\epsilon (q) =1+ 1/\alpha \gg 1$,
so that the TFA is valid for $B\gtrsim 1\,$T, i.e. $l_m \lesssim 30\,$nm.
Since in the linear screening regime the minimum value of the electron density
is $n_{\rm el}(0)=D_0 (E_F-V^q)$, linear screening breaks down if the
modulation strength becomes so large that $V^q \geq E_F$, i.e., for $V^q_{\rm
  ext} \geq \epsilon(q) \, E_F$.
For larger modulation amplitude the redistribution of electrons is
hindered: while the electron density at the minimum of $V_{\rm ext}(x)$ still
increases, the electron density at the maximum of $V_{\rm ext}(x)$ cannot
decrease further. Instead the density minimum becomes broader.
This means that the electrons are depleted from  strips along the maxima of
$V_{\rm ext}(x)$ and the 2DEG breaks off into a system of parallel quasi
one-dimensional ribbons.  Thus, the imposed
harmonic modulation potential $V_{\rm ext}(x)$ now leads to an anharmonic 
density distribution and, therefore, an anharmonic screened
potential. \cite{Wulf88:162}
 Mathematically, Eq.~(\ref{hartree}) with (\ref{kernel-hom}) and 
Eq.~(\ref{linear-resp}) now represents a nonlinear integral equation that must
be solved numerically. In Fig.~\ref{fig:odd} we plot the total variance
${\rm Var}[V]=V(0)-V(a/2)$ as a function of the amplitude $V^q_{\rm ext}\equiv
V_0$ of the imposed modulation potential for several values of the magnetic
field. The result for $B=0$ and $T=0$ is shown as a thick solid line.
 In the linear screening regime, ${\rm Var}[V]\equiv 2V^q
=2 V_0 /\epsilon(q)$. As linear screening breaks down, a kink appears in
the line and the variance increases much faster than in the linear
regime. With increasing temperature this kink is rounded off, while the 
${\rm Var}[V]$-vs-$V_0$ curve as a whole is not much affected  (shown
for $k_BT/E_F=0.04$ by the open circles in
Fig.~\ref{fig:odd}).
Here and in the following we measure energies in units of the Fermi energy
$E_F=\bar{n}_{\rm el}/D_0$ (for GaAs with $\bar{n}_{\rm el}\approx 3\cdot
10^{11}\,$cm$^{-2}$, $E_F\approx 10\,$meV), and we keep the mean electron
density  $\bar{n}_{\rm el}$, and thus $E_F$, constant. We will focus in the
following on the regime $V_0 \lesssim \epsilon(q) E_F$, where screening is
linear in the limit $B=0$, $T=0$.

%

\subsubsection{Half filled Landau levels}

With the Landau DOS [see Eq.~(\ref{landau-dos})] and the definition
 of a position-dependent chemical potential, $\mu(x)=\mu^{\star}-V(x)$,
 Eq.~(\ref{thomas-fermi}) yields  
\be \label{landau-density}
n_{\rm el}(x)=\hbar \omega_c D_0 \sum_n f\big(E_n- \mu(x)\big) .
\ee
%
We may also write the argument of the Fermi function
 as $E_n(x)-\mu^{\star}$ and  interpret 
$E_n(x)=E_n+V(x)$ as  position-dependent Landau energies, which is correct if
the Thomas-Fermi approximation holds.
It will be useful to define, in addition to the average LL filling factor
$\bar{\nu}=2\pi l_m^2 \bar{n}_{\rm el}$, a local filling factor 
$\nu(x)=2\pi l_m^2 n_{\rm el}(x)$.

For $k_BT \ll \hbar \omega_c$ and a homogeneous 2DEG
with  partly filled $n$-th Landau level, $\mu\sim E_n$, $\bar{\nu} =
2n+\nu_n$, where $\nu_n \approx 2 f(E_n-\mu)$ is the filling factor of
the $n$-th Landau level,  the TDOS is \cite{Wulf88:4218}
\be \label{landau-TDOS}
 D_T(\mu;B) \equiv \frac{\partial \bar{n}_{\rm el}}{\partial \mu} =
 \frac{\hbar \omega_c}{k_BT}\,  \frac{\nu_n}{2}\left(1- \frac{\nu_n}{2}\right)
 D_0 \,,
\ee
 which is
 peaked around $\mu=E_n$ with a maximum value $ \hbar \omega_c D_0/(4k_BT)$
and a width of order $k_BT$ [a crude approximation is $D_T(\mu;B) \approx
( \hbar \omega_c D_0/4k_BT)\theta (2k_BT-|E_n- \mu|)$]. 
 Linearizing Eq.~(\ref{landau-density}) 
with respect to the screened potential $V(x)$,
we obtain the Eq.~(\ref{linear-eps}) with  $Q_0$
replaced by $Q_B= Q_0 D_T(\mu;B)/D_0 \gg Q_0$,
\be \label{epsilon_B}
 \epsilon(q;B)=1+ \frac{\hbar \omega_c}{k_BT}\,  \frac{\nu_n}{2}\left(1-
 \frac{\nu_n}{2}\right) \frac{Q_0}{|q|}\,.
\ee
 For exactly half filling,
 $\mu=E_n$, his is a rather good 
approximation, as can be seen from Fig.~\ref{fig:odd}, which shows numerical
results for $\bar{\nu}=\nu_0=1$ ($n=0$) at two different temperatures (two
 lowest 
 curves at $V_0/E_F >15$). We see that the screening at finite magnetic field
 depends much stronger on temperature than at $B=0$.
 The linear approximation breaks down, if the amplitude of the
screened potential becomes of the order $2k_BT$.

This yields, in the limit of low temperatures and for $\nu_n=1$, the estimate
  for the linear  screening regime, 
\be \label{estimate}
\frac{V^q_{\rm ext}}{E_F} \lesssim  \epsilon(q;B)\frac{2 k_BT}{E_F} \approx 
\frac{ \epsilon(q)}{\bar{\nu}} \,, 
\ee
with $ \epsilon(q)= \epsilon(q;B=0)=1+1/\alpha$.
 For a larger modulation the
redistribution of electrons within the considered LL is not efficient enough
to screen the imposed modulation potential, and similar to the $B=0$ case, the
 variance of the screened potential increases much stronger than in the 
linear regime.
 In Fig.~\ref{fig:odd} we show low-temperature ($k_BT/E_F=0.001$)
results for odd-integer $\bar{\nu}$ values, calculated numerically from
Eqs.~(\ref{hartree}), (\ref{kernel-hom}), and (\ref{landau-density}).  For
this temperature, the linear increase of the 
screened potential with the applied modulation amplitude $V_0$ is not resolved
on the scale of Fig.~\ref{fig:odd}. However the rapid increase of the variance
of the screened potential at $V_0/E_F \sim  \epsilon(q)/\bar{\nu}$ is clearly seen
for the indicated $\bar{\nu}$ values.

 For $\bar{\nu}=1$ the situation is very similar to
the $B=0$ case, apart from the fact that  screening in the linear regime is
much stronger (``perfect screening'', ``pinning of lowest LL to Fermi level'')
due to the higher DOS.
For $\bar{\nu} > 1$ new phenomena occur, which we will now discuss.

\subsection{Emergence of incompressible strips \label{sec:2-pinning}}

\subsubsection{Odd-integer filling factor  $\bar{\nu}$}

\begin{figure}
{\centering \includegraphics[width=0.95\linewidth]{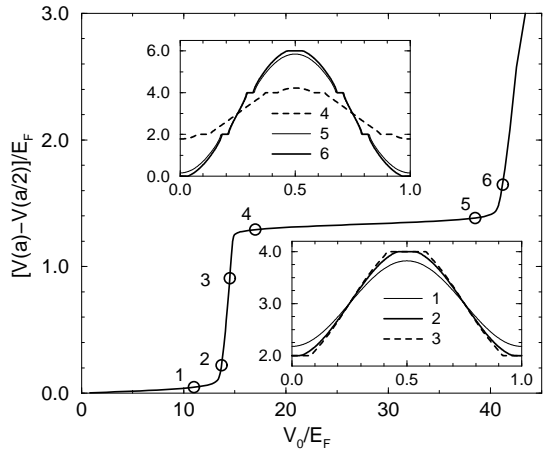}}
\caption{ \label{fig:dichte-nu3}
Variance of the total potential versus $V_0$, for average filling factor $\bar{\nu}=3$. The insets show the local
filling factor $\nu (x)$ in one modulation period ($0\leq x/a \leq 1$) for the
six $V_0$ values indicated by circles. Parameters:
 $k_BT/E_F=0.01$, $\epsilon(q)=41$, $q=2\pi/a$. }
\end{figure}

\begin{figure}
{\centering \includegraphics[width=\linewidth]{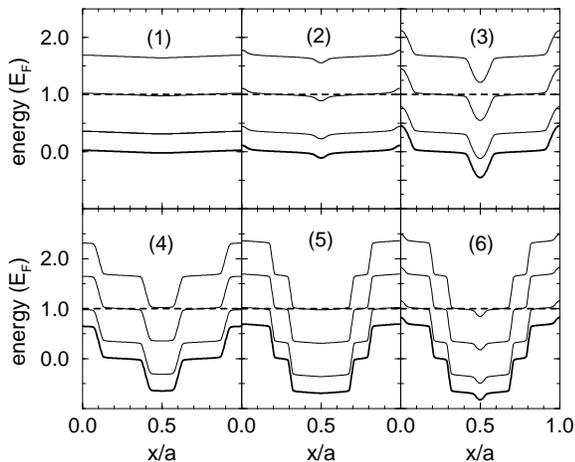}}
\caption{ \label{fig:pot-nu3}
Total potential (thick solid lines) and the three lowest of the corresponding
Landau levels (thin solid lines) together with the electrochemical potential
(thick dashes lines) for the six $V_0$ values indicated in
Fig.~\protect\ref{fig:dichte-nu3}. Parameters as in that figure. }
\end{figure}

We start with filling factor $\bar{\nu}=3$ and investigate the changes of the
electron density (Fig.~\ref{fig:dichte-nu3}) and of the  total potential
(Fig.~\ref{fig:pot-nu3})  with increasing  
amplitude $V_0$ of the imposed modulation $V_{\rm ext}(x)=V_0 \cos qx$, giving
explicit results for the six typical $V_0$ values indicated by open circles in
Fig.~\ref{fig:dichte-nu3}.

For  $V_0=11 \, E_F$, close to the breakdown of
linear screening (case 1), the density is strongly modulated (see thin line in
lower inset of Fig.~\ref{fig:dichte-nu3}), but the modulation appears still
cosine-like. The potential is so effectively screened that the second-lowest
LL ($n=1$) is pinned (within a few $k_BT$) to the Fermi energy [see
Fig.~\ref{fig:pot-nu3}(1)]. 
For case 2, $V_0=13.7\, E_F$, the total potential has developed
locally confined maxima and minima, while it remains rather flat in between
[Fig.~\ref{fig:pot-nu3}(2)].  Near these extrema $|E_1(x)-\mu^{\star}|$
 becomes so large that the LL $n=1$ is completely occupied (near $x=a/2$) or
 empty (near $x=0$), and incompressible strips with local filling
 factors $\nu(x)=4$ and $\nu(x)=2$ develop near the potential minima and
 maxima, respectively [see thick solid line in the lower inset of
 Fig.~\ref{fig:dichte-nu3}, and Fig.~\ref{fig:pot-nu3}(2)].  
Increasing the modulation to $V_0=14.5\,E_F$ (case 3) leads to more pronounced
local extrema and broader incompressible strips, but does not change the
situation qualitatively. The overall change of the density distribution is
rather small (see lower inset of Fig.~\ref{fig:dichte-nu3}), indicating poor
screening. Indeed the slope of the Var$[V]$-vs-$V_0$ curve in this regime is 
$\Delta {\rm Var} /\Delta V_0 \approx 1$, i.e., only slightly smaller than in
the absence of any screening, which would yield $\Delta {\rm Var} /\Delta V_0
=2$. We note that in the incompressible strips the local filling factor (i.e.,
the density) is constant, while in the pinning regions, i.e., the compressible
strips, the screened potential still has a finite slope, proportional to
$k_BT$. \cite{Lier94:7757}

As $V_0$ increases further to case 4 ($V_0=17\,E_F$), the modulation
becomes so strong that the maximum
of the lowest LL, $E_0(0)$, and the minimum of the lowest unoccupied LL,
$E_2(a/2)$, reach the Fermi level $\mu^{\star}$ (to within $k_BT$).  Then
thermal population of 
the higher LL ($n=2$ near $x=a/2$) and depletion of the lower LL ($n=0$ near
$x=0$) starts and compressible strips emerge in the center of each
incompressible strip [see dashed line in the upper inset of
Fig.~\ref{fig:dichte-nu3} and Fig.~\ref{fig:pot-nu3}(4)]. Further increase of
$V_0$ up to $V_0=38.5\,E_F$ (case 5) widens the compressible strips and leads
to a strong increase of the density modulation due to a redistribution of
electrons from the $n=0$ to the $n=2$ LL. This results in a strong screening,
similar to that in the linear screening regime at weak modulation, and, apart
from a weak increase with a slope proportional to $k_BT$, the variance  
$ {\rm Var}[V]= V(0)-V(a/2)$ remains constant at the value
$ {\rm Var}[V]= \mu^{\star}-E_0(0) - [ \mu^{\star}-E_2(a/2)]=2 \hbar \omega_c$.
This plateau behavior of the Var$[V]$-vs-$V_0$ curve is obviously an immediate
consequence of the pinning of LLs to the Fermi level, i.e., of the nearly
perfect screening.

In case 5 we reach a situation in which the lowest ($n=0$) LL is nearly empty
at the potential maximum and the higher ($n=2$) LL nearly full at the
potential minimum.
 In case 6 ($V_0=41.2\, E_F$) the total potential again develops local
 extrema, similar to the situation depicted in Fig.~\ref{fig:pot-nu3}(2).
But now the incompressible strip created at the potential maximum is due to
the depopulation of the lowest LL, i.e., due to vanishing electron density.
With further increasing $V_0$ the depletion regions become wider and the
density near the potential minima increases, but screening remains much poorer
than in the plateau region.

We see from Fig.~\ref{fig:dichte-nu3} that the global appearance of the
density modulation, apart from a fine-structure related to the
incompressible strips, is more or less cosine-like. 
We will use this finding for a rough estimate of the plateau width of the 
 Var$[V]$-vs-$V_0$ curves.
First we conclude from the cosine-like form of the induced density variation
that in the high-screening plateau region, along with Eqs.~(\ref{harmonic})
and (\ref{linear-eps}),  the first relation
of Eq.~(\ref{estimate}) holds qualitatively and relates the changes $\delta V
$ of the total potential to the changes $ \delta V_0 $ of the externally
applied potential by  $\delta V \sim
\epsilon(q;B) \delta V_0$. For $Q/|q| \gg 1$, this yields for the change of the
variance ${\rm Var}[V]$ across the plateau of width $\Delta V_0$
\be \label{plateau-slope}
\Delta {\rm Var}[V] \sim \frac{8 k_BT}{\hbar \omega_c \epsilon(q)} \Delta V_0
\,,
\ee
i.e., an estimate for the slope of the Var$[V]$-vs-$V_0$ curve in the plateau
 region. Since the modulation induces density changes $\delta n_{\rm
 el}$ mainly within the compressible regions of high TDOS, we estimate
$\delta n_{\rm el} \sim - D_T(\mu;B) \delta V \sim - (\hbar \omega_c/4 k_BT)
D_0\delta V $  (which holds for $\nu_n=1$ and $|\delta V| \lesssim 2k_BT$). In
 terms of $\delta \nu =2\pi l^2_m  \delta n_{\rm el}$ this 
yields $\delta V \sim 2 k_BT \delta \nu$, and together with
 Eq.~(\ref{plateau-slope}) the relation
\be \label{plateau-width}
 \quad  \frac{\Delta V_0}{E_F}  \sim \frac{\epsilon(q)}{2 \bar{\nu}} 
\Delta {\rm Var}[\nu] 
\ee
between  the plateau width $\Delta V_0$ and the change of the filling factor
variance ${\rm Var}[\nu]=\nu(a/2)-\nu(0)$ (defined at fixed $V_0$) across the
plateau. 
This criterion applies also to the small-$V_0$ linear-screening regime, in
which $\nu(x)$ varies within the same LL, with $ {\rm Var}[\nu]$ increasing
from 0 to 2, i.e.\  $\Delta {\rm Var}[\nu] =2$ [see Eq.~(\ref{estimate})]. The
resulting width of the  linear-screening regime, $\Delta V_0/E_F 
\approx \epsilon(q) / \bar{\nu}$, describes the numerical results of
Fig.~\ref{fig:odd} for odd-integer $ \bar{\nu}=2n+1$ quite well.
From the discussion of Figs.~\ref{fig:dichte-nu3} and \ref{fig:pot-nu3} we
expect that, for $n>0$ the linear regime of the ${\rm  Var}[V](V_0)$ curve
 is terminated by a step  of height $2 \hbar \omega_c = 4E_F/  \bar{\nu}$,
which is followed by a plateau. While $V_0$ sweeps through the plateau,
in addition to the LL with index $n$  the two LLs with indices $n-1$ and $n+1$
are locally pinned to the Fermi level and lead to a total change $\Delta  {\rm
  Var}[\nu]=4$ [for  $\bar{\nu}=3$ from ${\rm Var}[\nu]=\nu(a/2)-\nu(0)=2$ on
the left side to $ {\rm  Var}[\nu]=6$ on the right side of the plateau, as
seen from the upper inset of Fig.~\ref{fig:dichte-nu3}]. This yields the
plateau width  $\Delta V_0/E_F \approx 2\epsilon(q) /\bar{\nu}$. 
If $n-1=0$, the plateau will be followed by the breakdown regime. If $n-1>0$,
the plateau will be followed by a further step of the same height to a plateau
of the same width.

To summarize: at very low temperatures and odd-integer filling factors $
\bar{\nu}=2n+1$,  the variance ${\rm  Var}[V]$
of the screened potential as function of the imposed modulation amplitude 
 $V_0$ shows a linear screening regime for  $V_0 \lesssim
\epsilon(q) E_F/\bar{\nu}$ which is followed by $n$ successive steps of
height $2 \hbar \omega_c = 4E_F/  \bar{\nu}$ and width $\Delta
V_0 \approx 2\epsilon(q) E_F /\bar{\nu}$. 
The plateau of the $n$-th step ends at the breakdown of the 2DEG into a
pattern of isolated 1D systems, which leads to poor screening and is indicated
in the ${\rm  Var}[V](V_0)$ curve by a slope of order unity. At finite
temperature, the plateaus assume finite positive slopes,
 which are estimated from Eq.~(\ref{plateau-slope}) as
$\Delta {\rm  Var}[V] /\Delta V_0 \approx 4 \bar{\nu} k_BT/[ \epsilon(q)
E_F]$.
These results, which describe the content of the numerically calculated
Figs.~\ref{fig:odd}-\ref{fig:pot-nu3} very well, depend, of course, on the high
symmetry of the situations considered so far.


\subsubsection{Even-integer filling factor $\bar{\nu}$}

Another situation of high symmetry is that of an even-integer filling factor 
$\bar{\nu}=2n+2$, where $n$ is the index of the highest occupied LL
and the  Fermi energy $E_F=\hbar \omega_c (n+1)$ lies in the middle between two
adjacent LLs. According to  Eqs.~(\ref{estimate}) and (\ref{epsilon_B}), the
linear screening 
regime shrinks to zero, since $\nu_n=2$. Thus, at very low  temperature
($k_BT \ll \hbar \omega_c/2$) a weak modulation $V_{\rm ext}(x)=V_0
\cos qx$ will not be screened, 
i.e., the local filling factor will be independent of the modulation, $\nu(x)
\equiv \bar{\nu}$, and the total potential will equal the external one, with
variance ${\rm Var}[V](V_0)=2 V_0$. This situation changes when the modulation
potential becomes so large that $|V_0 - \hbar \omega_c/2| \sim  k_BT$,
i.e., the  maximum energy $E_n(0)$ of the highest occupied, and the minimum
energy 
$E_{n+1}(a/2)$ of the lowest unoccupied LL approach the Fermi energy. Then,
with increasing $V_0$ the LL $n$ is depleted near $x=0$ while the LL $n+1$ is
populated near $x=a/2$, forming compressible strips with local filling factors
$\nu(x)<\bar{\nu}$ and $\nu(x)>\bar{\nu}$, respectively. 

\begin{figure}
{\centering \includegraphics[width=\linewidth]{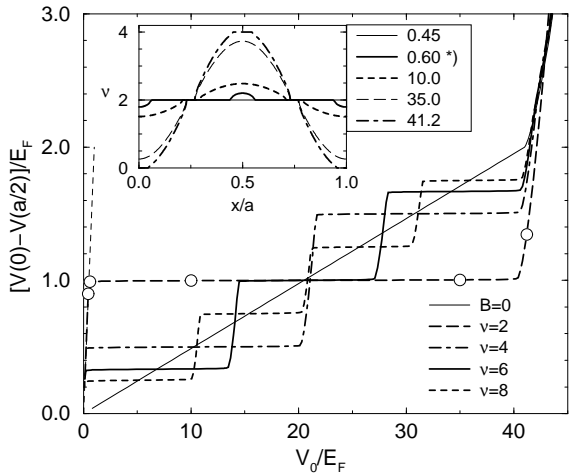}}
\caption{ \label{fig:even}
As in Fig.~\protect\ref{fig:odd} but for
 even-integer values of the (average) filling factor ($k_BT/E_F=0.001$,
 $\epsilon(q)=41$). The thin solid line indicates the result for $B=0$, $T=0$,
 the thin dashed line has slope 2. The inset shows the local filling factor
 $\nu(x)$ in one modulation period for average filling $\bar{\nu}=2$ and the
 five values of $V_0$ indicated by circles in the main figure. *) For
 $V_0/E_F=0.60$ the  deviation $[\nu(x)-2]$ is enhanced by a factor 20. }
\end{figure}

This is demonstrated in the inset of Fig.~\ref{fig:even}, which shows the
local filling factor $\nu(x)$ for $k_BT/E_F=0.001$ and the
average filling factor $\bar{\nu}=2$, i.e., $\hbar \omega_c=E_F$,
and for the modulation strengths $V_0$ indicated by circles
in the main figure. For  $V_0/E_F=0.45$ in the non-screening region the
deviation $[\nu(x)-\bar{\nu}]$ is practically zero (numerically 
$<10^{-6}$), while for $V_0/E_F=0.6$ it is finite, although small (in the inset
enhanced by a factor of 20), with narrow compressible strips. Between
$V_0/E_F=0.5$ and $V_0/E_F\approx 41$ the width of the compressible strips and
the deviation $[\nu(x)-\bar{\nu}]$ increase continuously, while the variance
${\rm Var}[V] \approx \hbar \omega_c=E_F$ remains constant.
Since screening is due to the redistribution of electrons at the Fermi energy,
i.e., to electrons in
the compressible strips where the TDOS is large, we may again use
Eq.~(\ref{plateau-width}) to estimate the plateau width. At the beginning of
the first plateau the filling factor  is constant, $\nu(x) \equiv \bar{\nu}$,
i.e., ${\rm   Var}[\nu]=0$. At the end of 
the plateau, the LL $n$ is depleted at the potential maximum, $\nu(0)=
\bar{\nu}-2$, and at the potential minimum the LL $n+1$ is full,
$\nu(a/2)=\bar{\nu}+2$, i.e.,  ${\rm  Var}[\nu]=4$.
 Thus, we have to use Eq.~(\ref{plateau-width}) with
$\Delta {\rm  Var}[\nu] =4$ and obtain for the plateau width $\Delta V_0 \sim 2
\epsilon(q) E_F/\bar{\nu}$. 

This estimate is obviously in good agreement with the numerical
calculations presented in Fig.~\ref{fig:even}.
For filling factor $\bar{\nu}=2$, i.e., $n=0$, the first plateau ends at the
transition to the poor-screening breakdown regime, since then the lowest LL
$n=0$ is completely depleted at the potential maxima (thick dash-dotted line
in inset of Fig.~\ref{fig:even}).
 For  $\bar{\nu}=2n+2$ with $n>0$  a behavior similar to that discussed
in Fig.~\ref{fig:pot-nu3} occurs. As $V_0$ increases slightly beyond the
plateau regime, a narrow local maximum of $V(x)$ develops near $x=0$,
accompanied with an incompressible strip of filling $\nu(x)=2n$ due to
the local depletion of the LL $n$. Simultaneously, a  narrow local minimum of
$V(x)$ develops near $x=a/2$, accompanied with an incompressible strip of
 filling $\nu(x)=2n+4$ due to the local occupation of the LL $n+1$. Then,
in a narrow $V_0$ interval these new extrema become more pronounced and the
accompanied incompressible strips widen a little. However, the accompanied
density change is small, resulting in a poor screening and a rapid increase of
the  ${\rm Var}[V](V_0)$ curve. This interval ends when the new maximum
$E_{n-1}(0)$ of the Landau level $n-1$ and the new minimum $E_{n+2}(a/2)$ of
the LL $n+2$ come close to the Fermi level (within a few $k_BT$).
Then, with further increasing $V_0$, new compressible strips open at the
locations of the potential extrema, and a plateau region of the 
${\rm Var}[V](V_0)$ curve with ``perfect'' screening sets in.
We thus again find a step behavior like in Fig.~\ref{fig:odd} with step height
$\Delta {\rm Var}[V]=2 \hbar \omega_c =4E_F /\bar{\nu}$. During the $V_0$
sweep through the corresponding plateau the LL $n-1$ will be depleted near
$x_0$ while the LL $n+2$ is occupied near $x=a/2$. Thus, we can estimate the
plateau width from Eq.~(\ref{plateau-width}) with $\Delta  {\rm Var}[\nu]
=4$. The last plateau is the one corresponding to the local depletion of the
$n=0$ LL. 

In summary, for $\bar{\nu}=2n+2$ and very low temperature, the ${\rm
  Var}[V](V_0)$ curve shows a linear increase with slope 2 for $0 \leq V_0
  \leq E_F/ \bar{\nu}$, followed by a plateau of height $\hbar \omega_c= 2E_F/
  \bar{\nu}$ and width $\Delta V_0 \sim  2\epsilon(q) E_F/\bar{\nu}$. 
This plateau is followed by $n$ steps of height $ 2\hbar \omega_c$
and approximately the same width $\Delta V_0$. The plateau of the last step is
  followed by the breakdown regime.


\subsubsection{Non-integer filling factor $\bar{\nu}$}

\begin{figure}
{\centering \includegraphics[width=\linewidth]{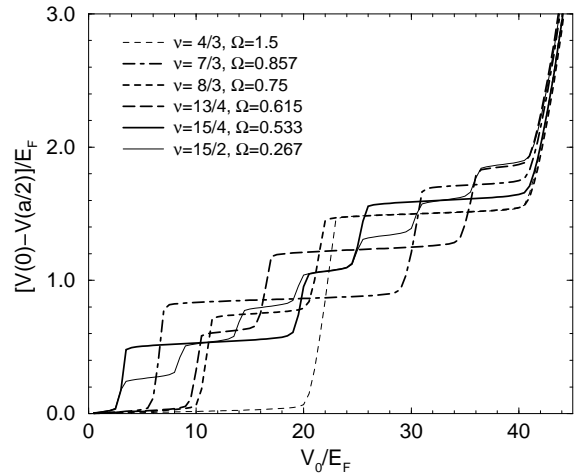}}
\caption{ \label{fig:nonint}
As in Fig.~\protect\ref{fig:odd} but for some non-integer
  values of the (average) filling factor and for higher temperature,
  $k_BT/E_F=0.01$ ( $\epsilon(q)=41$). }
\end{figure}

In Fig.~\ref{fig:nonint} we show ${\rm Var}[V](V_0)$ curves for a few
non-integer values of the average filling factor $\bar{\nu}=2n +\nu_n$, with
$0< \nu_n <2$. Although these results may, at a first glance, look confusing,
we will now demonstrate, that they can easily be understood, and even
predicted, from a few simple principles.

To estimate the width of the linear screening regime at small $V_0$ values, 
we follow the reasoning of Sec.~\ref{sec:2-breakdown}, 
$V^q_{\rm ext} \approx  \epsilon(q;B) V^q$, but we note that the linear
approximation to the Taylor expansion of $n_{\rm el}(x)$ with respect to
$V(x)$ [see Eq.~(\ref{landau-density})] 
does no longer hold for $|V(x)|\sim 2 k_BT$, since for $\nu_n \neq 1$ the
second order term [$\propto \partial^2 n_{\rm el}/\partial \mu^2 =
(\partial n_{\rm el}/\partial \mu)(1-\nu_n)/k_BT$] yields already noticeable
contributions for smaller $V(x)$. To take this into account, we use the linear
approximation only for $|V(x)|\lesssim 2 k_BT/(1+|1-\nu_n|)$ and obtain as
condition for the linear screening regime
\be \label{estimate-nonint}
\frac{V_0}{E_F} \lesssim \frac{\epsilon(q)}{\bar{\nu}}\, \frac{\nu_n
  (2-\nu_n)}{1+|1-\nu_n|}\,.
\ee
For $\nu_n =1$ this reduces to the estimate (\ref{estimate}). But in addition,
Eq.~(\ref{estimate-nonint}) states
that for even-integer filling, $\nu_n \rightarrow 0$ or $\nu_n \rightarrow 2$,
the linear screening regime shrinks to zero, and it provides a good
description of the widths of the linear screening regimes in the examples shown
in Fig.~\ref{fig:nonint}.

We will now use Eq.~(\ref{plateau-width}) to obtain estimates of the
plateau widths and heights of the ${\rm Var}[V]$-vs-$V_0$ curves, which
contain the estimate (\ref{estimate-nonint}) for the linear screening regime
as a special case. Our estimates are based on the observation that in all cases
we have studied the density modulation is nearly symmetric about the average
density, so that the average of extreme values of the local filling factor is
close to the average filling, $\nu(0) +\nu(a/2) \approx 2 \bar{\nu}$. Nearly
perfect screening occurs, if at both the potential maxima and  the minima a
LL is pinned to the Fermi level, so that electrons can easily be
redistributed between these LLs and both $\nu(0)$ and $\nu(a/2)$ are different
from even integers. If, with increasing $V_0$, $\nu(0)$ approaches an even
integer value, $\nu(0)=2k$, the LL $k$ is depleted at $x=0$ and a local
maximum of $V(x)$ starts to develop there. Screening remains poor, and ${\rm
  Var}[V](V_0)$ increases rapidly,  until the LL
$k-1$ reaches the Fermi level at $x=0$. Then a step of height $\hbar \omega_c$
is completed and the next plateau with perfect screening starts. A similar
step begins as $\nu(a/2)$ reaches the value $2 k'$. Then the LL $k'-1$ is
completely filled at $x=a/2$ and a local potential minimum starts to develop
there. Perfect screening begins again, if the LL $k'$ reaches Fermi level at
$x=a/2$ and the increase of ${\rm  Var}[V](V_0)$ by $\hbar \omega_c$ is
completed.

 We combine now these consideration with the estimate (\ref{plateau-width}).
For convenience, we introduce the dimensionless variables
\be \label{dimensionless}
v=\frac{V}{E_F} \quad v_0=\frac{V_0}{\epsilon(q) E_F}\,, \quad
\Omega=\frac{\hbar\omega_c}{E_F}\,,
\ee
and focus on the regime $0<v_0<1$, in which
for $T=0$ and $B=0$ screening is linear and leads to ${\rm  Var}[v](v_0)=2
v_0$. 
To keep the discussion simple, we consider the two possible cases of
non-integer $\bar{\nu}=2n+\nu_n$ separately.

\paragraph{For $0<\nu_n <1$,} the end of the linear screening
region, where $v\ll \Omega$, is reached when $\nu(0)=2n$. Then
$\nu(a/2)\approx \bar{\nu}+ \nu_n$  and ${\rm  Var}[\nu]=2\nu_n$, and across
the linear screening regime we find $\Delta {\rm  Var}[\nu]=2\nu_n$. According
to Eq.~(\ref{plateau-width}), linear screening ends at $V_0/E_F \sim
\epsilon(q) \nu_n/\bar{\nu}$, in agreement with Eq.~(\ref{estimate-nonint}).
Neglecting potential variations $\propto k_BT \ll \Omega$, we note
\be \label{eq:linscr-1}
{\rm  Var}[v] \approx 0\,, \quad \mbox{if} \quad 0 < v_0 < 1-n \Omega \,,
\ee
since $\nu_n=\bar{\nu}-2n$ and $\Omega=2/\bar{\nu}$. If $n=0$, larger $V_0$
lead to the poor-screening quasi 1D ribbon regime.

For $n>0$, the next plateau terminates when $\nu(a/2)=2n+2$. Then
$\nu(0)\approx 2n- 2(1-\nu_n)$, i.e., ${\rm  Var}[\nu]=2(2-\nu_n)$. Across
this plateau we have    $\Delta {\rm  Var}[\nu]=4(1-\nu_n)$, and
with Eq.~(\ref{plateau-width}) 
$\Delta V_0 \approx 2 \epsilon(q) E_F (1-\nu_n)/\bar{\nu}$. This yields
$\Delta v_0 \approx (2n+1)\Omega-2$ and
\be  \label{eq:plateau1-1}
{\rm  Var}[v] \approx \Omega\,, \quad \mbox{if} \quad  1-n \Omega 
<v_0 < (n+1)  \Omega-1 \,.
\ee
This plateau is followed by another one along which $\nu(0)$ decreases to
$2(n-1)$, while $\nu(a/2)$ increases to $\approx 2n+2+2\nu_n$ and thus
 ${\rm  Var}[\nu]$ to $4+2\nu_n$. Thus, across that plateau we find
$\Delta {\rm  Var}[\nu]=4\nu_n$ and $\Delta v_0= 2-2n \Omega$, which leads to
\be  \label{eq:plateau2-1}
{\rm  Var}[v] \approx 2\Omega\,, \; \; \mbox{if} \; \;
 (n\!+\!1)  \Omega\!-\!1 < v_0 < 1\!-\!(n\!-\!1)\Omega \,.
\ee
If $n=1$, this plateau is followed by the poor screening
 quasi 1D ribbon regime.
 If $n>1$, we are in the same situation as at the end of
the low-$V_0$ linear screening regime, and a double step of total width
$\Omega$, consisting of one step of height $\Omega$ and
plateau width $\Delta v_0=(2n+1)\Omega-2$ and another one of  height $\Omega$
and width $\Delta v_0 =2-2n\Omega$, will follow.  Thus, we obtain for $0<k\leq
n$ 
\be  \label{eq:plateau_k1-1}
{\rm Var}[v] &\approx& (2k-1) \Omega\,,  \quad  \mbox{if}\\
&~& 1-(n+1-k) \Omega <v_0 < (n+k) \Omega-1\,, \nonumber\\
{\rm Var}[v] &\approx& 2k \Omega\,,  \quad  \mbox{if} \label{eq:plateau_k1-2}
\\ 
&~& (n+k) \Omega-1 <v_0 <1-(n-k) \Omega\,. \nonumber
\ee
Thus, the linear screening regime
is followed by $n$ double steps, which sum up to a total width $\Delta v_0=1$,
and on the last plateau  (before breakdown) we have ${\rm
  Var}[v]=2(1-\nu_n/\bar{\nu})$.  For
$0<\nu_n <0.5$, the first plateau of the double step is wider than the second
one, as for the dash-dotted line in Fig.~\ref{fig:nonint} ($\bar{\nu}=2.33$),
while for $0.5 <\nu_n <1$ the second plateau of the double step is the wider
one, as for the short-dashed line ($\bar{\nu}=2.67$).

We should mention three limits. For $\nu_n \rightarrow 0$ the low-$V_0$ linear
screening regime shrinks to zero and the first plateau of the double step
exhausts its full width, so that the second step merges with the first one of
the following double step. Thus, we observe at small $V_0$ a step of height
$\Omega$, followed by steps of the double height $2\Omega$,
and all plateaus have the same widths, as we found previously. For $\nu_n=0.5$
we get an even number of steps
which all have the same heights and widths.  For $\nu_n \rightarrow 1$ the
width of the first plateau of each double step shrinks to zero, so that the 
low-$V_0$ linear screening regime is followed by steps of height $2
\Omega$ and width $\Omega$, as we have seen before.

\paragraph{For $1<\nu_n <2$,} we have at the end of the linear screening
regime  $\nu(a/2)=2n+2$ and $\nu(0)\approx 2n+2(\nu_n-1)>2n$, with ${\rm
 Var}[\nu]=2(2-\nu_n)$. From Eq.~(\ref{plateau-width}) we obtain
\be \label{eq:linscr-2}
{\rm  Var}[v] \approx 0\,, \quad \mbox{if} \quad 0 < v_0 < (n+1) \Omega -1\,,
\ee
and see that now the linear screening regime  is always followed by a step of
height $\Omega$ to a plateau of perfect screening. For $n=0$ (i.e.,
$1<\Omega<2$) this plateau covers
the interval $\Omega-1<v_0<1$  (see e.g., thin dashed
line of Fig.~\ref{fig:nonint} for $\bar{\nu}=1.33$).
  To estimate for $n>0$
height and width of the following steps and plateaus,   respectively, 
  we proceed as before, exploiting that at the end of each 
plateau either $\nu(0)$ or $\nu(a/2)$ reaches an even integer value, and that 
$\nu(0)+\nu(a/2) \approx 2 \bar{\nu}$. The result is, for $0< k \leq n$,
a double step of total width $\Omega$,
\be  \label{eq:plateau_k2-1}
{\rm Var}[v] &\approx& (2k-1) \Omega\,,  \quad  \mbox{if}\\
&~& (n+k) \Omega-1 <v_0 <1- (n+1-k) \Omega\,, \nonumber\\
{\rm Var}[v] &\approx& 2k \Omega\,,  \quad  \mbox{if} \label{eq:plateau_k2-2}
\\ 
&~& 1-(n+1-k) \Omega <v_0 <(n+k+1) \Omega-1\,, \nonumber
\ee
which is followed by a final single step, 
\be  \label{eq:plateau_last-2}
{\rm Var}[v] \approx (2n+1)  \Omega\,, \;\;  \mbox{if}\;\;
 (2n+1)  \Omega -1 < v_0 <1\,,
\ee
of height $\Omega$ and plateau width $2-(2n+1)  \Omega$. The linear screening
regime [width $(n+1) \Omega-1$], the $n$ double steps and this final plateau
cover together the interval $0<v_0<1$, as in the case $0<\nu_n<1$. The
variance of the screened potential in the last plateau is ${\rm
  Var}[v]=2[1-(\nu_n-1)/\bar{\nu}]$. 

For $1<\nu_n<1.5$ the first plateau of each double step is wider than the
other plateaus, as seen in Fig.~\ref{fig:nonint} for the long-dashed line
($\bar{\nu}= 3.25$), which exhibits one double step following the initial
single step. For $1.5 <\nu_n <2$ these first plateaus are the
narrower ones, as seen for the thick solid line ($\bar{\nu}= 3.75$).
For $\nu_n=1.5$, the double steps consist of two individual steps of equal
heights and plateau widths, as is illustrated by the thin solid line ($\bar{\nu}=7.5$).

In the limit of odd-integer $\bar{\nu}$, $\nu_n \rightarrow 1$, the width of
the first single step together with the second plateau width of each double
step shrinks to zero, so that only steps with step height $2 \Omega$
and plateau width $\Omega$ occur. (For $\bar{\nu}=1$,
i.e., $n=0$, no double step exists and the single step merges with the
breakdown regime.) In the even-$\bar{\nu}$ limit, $\nu_n \rightarrow 2$, the
width of the first plateau of each double step shrinks to zero. Thus, the
first step of width $\Omega$ and height $ \Omega$
is followed by $n$ steps of the same plateau width but double step height.

\begin{figure}[h]
{\centering \includegraphics[width=\linewidth]{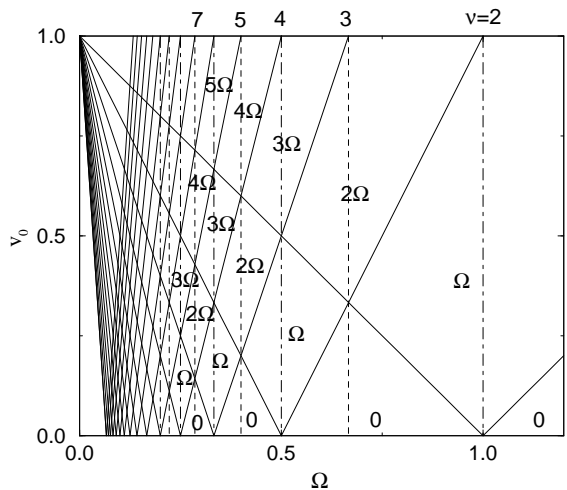}}
\caption{ \label{fig:slines}
The solid lines indicate $v_0=k\Omega-1$ and $v_0=1-k\Omega$ for $v_0=V_0/[E_F
\epsilon(q)]$, $\Omega=\hbar \omega_c/E_F$, and $k=1,\,2,\dots,\,15$. Even
(odd) integer values of the average filling factor are indicated by
dash-dotted (dashed) vertical 
lines. Within each area defined by the solid lines the value of ${\rm
  Var}[V]/E_F$ equals an integer multiple of $\Omega$. This value increases by
$\Omega$, if a solid line is crossed in upward direction. 
The region $v_0>1$ corresponds to the poor screening regime of parallel,
disconnected quasi 1D electron systems. }
\end{figure}
\paragraph{Summarizing} the estimates of this Sec.~\ref{sec:2-pinning}, we
note that the Eqs.~(\ref{eq:linscr-1}) -- (\ref{eq:plateau_last-2}) define a
set of straight lines in the $v_0$-$\Omega$ plane, which separate areas in
each of which the variance ${\rm Var}[v](v_0)$ equals an integer multiple of 
$\Omega$. This is schematically shown in Fig.~\ref{fig:slines}. Position and
height of the steps of the ${\rm Var}[v](v_0)$ curve 
for a given value of $\bar{\nu}$ can immediately be read
off from this figure along the vertical line at $\Omega=2/\bar{\nu}$.


\subsection{Sweeping the magnetic field \label{sec:sweep-B}}

\begin{figure}[h]
{\centering \includegraphics[width=\linewidth]{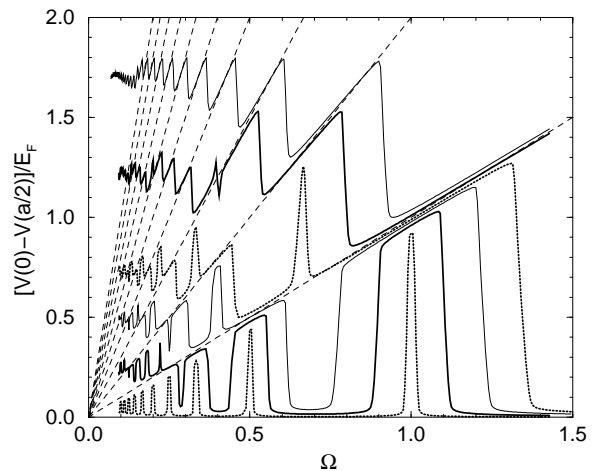}}
\caption{ \label{fig:varvsom}
Variance of the screened potential versus $\Omega_c=\hbar \omega_c/E_F$
for $V_0/E_F=1,\,5,\,10,\, 15,\,25, \,35$ (from bottom to top). The straight
dashed lines indicate integer multiples of the cyclotron energy.
 ($k_BT/E_F=0.01$, $\epsilon(q)=41$). }
\end{figure}

We now consider the screening of an external cosine potential $V_{\rm
  ext}(x)=V_0 \cos qx$ of fixed amplitude $V_0$ as function of the magnetic
  field $B$, keeping the average electron density at the fixed value of the
  positive background charge density. Then, with increasing $B$ the average
  filling factor $\bar{\nu}=2 E_F/\hbar \omega_c$ decreases. For the
  unmodulated 2DEG ($V_0=0$), this leads to the well known saw-tooth behavior
  of the chemical potential, which at low temperatures is pinned to the LLs,
  i.e.\ follows
  half-integer multiples of the cyclotron energy,
\be \label{sawtooth-mu}
\mu^{\star}=\hbar \omega_c (n+1/2)\quad \mbox{if} \quad
1/(n+1)< \Omega < 1/n \,.
\ee
In the modulated 2DEG ($V_0>0$), pinning of LLs to the electrochemical
potential causes  the total variance of the screened
potential to be an integer multiple of the cyclotron energy. Thus, for the
variance ${\rm Var}[V]$ as function of $B$ we expect a similar saw-tooth
behavior as for the $\mu^{\star}$-vs-$B$ curve. Numerical results for 
several values of $V_0$ are shown in Fig.~\ref{fig:varvsom}.
The uppermost curve for the largest modulation amplitude looks indeed similar
to a $\mu^{\star}$-vs-$B$ curve. However, whereas the latter with decreasing
$B$ always jumps to the next higher LL, the ${\rm Var}[V]$-vs-$B$ curves can
also jump back to the next lower LL, as is more clearly seen for the curves
with smaller modulation amplitudes. This seemingly irregular behavior of the 
${\rm Var}[V](\Omega)$ curves in Fig.~\ref{fig:varvsom} can easily be
understood from Fig.~\ref{fig:slines}, where we now have to follow horizontal
lines.
 For fixed
$v_0^0=V_0/[E_F \epsilon(q)]$ and decreasing $\Omega$, the variance ${\rm
  Var}[V]/E_F$ increases by $\Omega$, if the horizontal line  $v_0=v_0^0$
intersects one of the straight lines $v_0=k \Omega-1$, and decreases by 
$\Omega$, if it intersects one of the straight lines $v_0=1-k \Omega$.
For large $V_0$ ($v_0>0.6$), the variance jumps with decreasing $B$
monotonically to higher multiples of the cyclotron energy, until $B$ becomes
so small, that $\Omega=1-v_0^0$. Then, for smaller $B$ also jumps back to
lower multiples occur.
For small modulation amplitude ($v_0^0 \lesssim 0.1$) one has perfect
screening if the average filling factor is not too close to an even
integer. Near 
such values linear screening breaks down and ${\rm  Var}[V]/E_F$ approaches 
$\Omega$, provided $2V_0 > \hbar \omega_c$ (otherwise ${\rm  Var}[V]=2V_0$).
For sufficiently small $B$, of course the variance will equal higher multiples
of the cyclotron energy, so that the correct linear screening limit is
obtained in the limit of zero magnetic field.
Thus, if one adds the smoothening effect of finite temperature, one can
understand all properties of the apparently irregular ${\rm  Var}[V](\Omega)$
traces in  Fig.~\ref{fig:varvsom} in terms of the peculiar but regular
$v_0$-vs-$\Omega$ pattern sketched in Fig.~\ref{fig:slines}.

\section{Hall bar geometry}

\subsection{Boundary conditions and kernels}
We now consider a 2DEG with lateral confinement in the $x$ direction and
translation invariance in the $y$ direction, i.e., an idealized Hall bar
geometry. To study boundary effects on the screening properties, we will apply
an additional periodic external modulation potential in $x$ direction.
We will consider two different sets of boundary conditions, which lead to
slightly different confinement potentials.

\subsubsection{In-plane gates}

Following
Refs.~\onlinecite{Chklovskii92:4026,Chklovskii93:12605,Lier94:7757,Oh97:13519}
we first assume that all charges reside in the plane $z=0$, and that the
halfplanes $z=0,\,x<-d$ and $z=0, \, x>d$ are kept at constant electrostatic 
potential, $V(x,y,z=0)=0$ for $|x|>d$  (in-plane
gates). \cite{Chklovskii93:12605,Oh97:13519} Then the electrostatics can be
solved using the theory of complex functions, and the kernel in
Eq.~(\ref{hartree}), with $-x_l=x_r=d$, is obtained as \cite{Oh97:13519} 
\be
K_{\parallel}(x,t)=\ln
\left|\frac{\sqrt{(d^2-x^2)(d^2-t^2)}+d^2-tx}{(x-t)d} \right| \,.
  \label{kernel-inplane}
\ee
 Positive background charges
of the 2D charge density $e n_0$ between the in-plane gates will produce the
confinement potential (written as potential energy of an electron)
\be
V_{\rm bg}(x)=-E_0 \sqrt{1-(x/d)^2}\,, \quad E_0=2\pi e^2 n_0 d/\bar{\kappa}
\, , \label{vbg-inplane}
\ee
which can be calculated from Eq.~(\ref{hartree}) using the kernel
(\ref{kernel-inplane}) and replacing $n_{\rm el}(x')$ by $-n_0$.

\subsubsection{Perpendicular gates}

\begin{figure}
{\centering \includegraphics[width=0.8\linewidth]{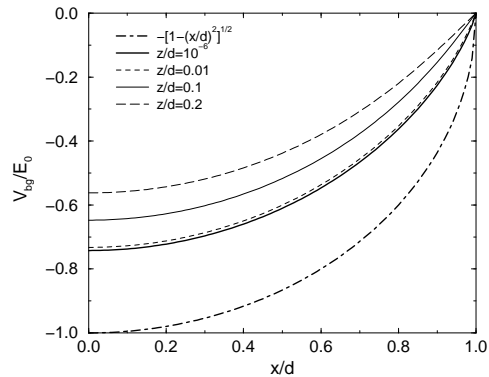}}
\caption{ \label{fig:confpot}
Confinement potential $V_{\rm bg}(x,y,z=0)/E_0$ due to a homogeneous plane of
 charge density $e n_0$ at distance $z$. The dash-dotted line is obtained from
model (\ref{kernel-inplane}) with $z=0$, the other lines are for model
 (\ref{perp-gates}). 
 }
\end{figure}

Another simple set of boundary conditions is obtained assuming the 2DEG to be
laterally confined by two equipotential planes located at $x=\pm d$ parallel
to the $y$-$z$ plane, $V(x=\pm d,y,z)=0$. This is a reasonable model for a free
standing mesa-etched Hall bar with free or metallized surfaces at $x=\pm d$,
which accommodate a large number of (partially occupied) surface states.
The electrostatics with these boundary conditions is well
known.\cite{Morse-Feshbach53:1240} 
In our notation it is expressed by the kernel
\be  \label{perp-gates}
K_{\perp}(x,t)=-\ln\! \left(\frac{\cos ^2 \frac{\pi}{4d}(x+t) + \gamma ^2}{
\sin ^2 \frac{\pi}{4d} (x-t) + \gamma ^2}\right)
\ee
for $\gamma \rightarrow 0$. Inserting this with
$\gamma=\sinh (\pi z/4d)$ into Eq.~(\ref{hartree}), where $-x_l=x_r=d$,
 yields the electrostatic
potential $V_H(x,y,z)$ due to the 2DEG at a position separated by the distance
$|z|$ from the plane of the 2DEG. Correspondingly, we can use this to
calculate the confinement potential produced in the plane of the 2DEG by a
plane, positive background charge at a distance $z$ from the 2DEG. Typical
confinement potentials are shown in Fig.~\ref{fig:confpot}. For $\gamma=0$ the
potential minimum is $V_{\rm bg}(0,y,0)/E_0=- 8 G/\pi ^2=- 0.74246$, with
Catalan's constant \cite{Gradshteyn} $G=0.915965594$.
\begin{figure}[h]
{\centering \includegraphics[width=\linewidth]{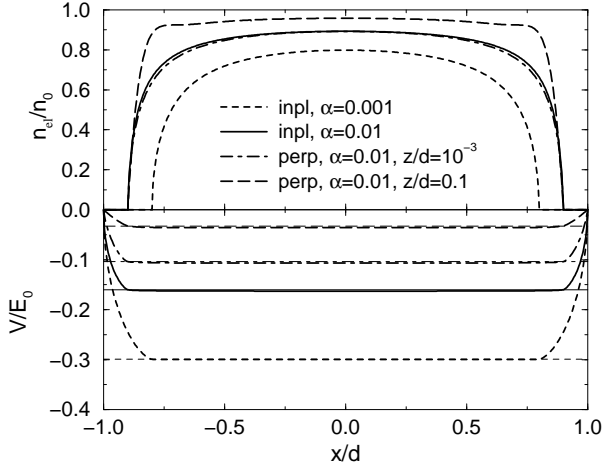}}
\caption{ \label{fig:denspot}
Some consistent density profiles (upper panel) and potentials (lower panel)
calculated for the in-plane-gates model  (\ref{kernel-inplane}) and the
perpendicular-gates model  (\ref{perp-gates}), respectively. The depletion
length is chosen as $d/5$ for the short-dashed curves and as $d/10$ else.
 Thin horizontal
lines indicate the corresponding electrochemical potentials. $\alpha=\pi
a_B^{\star}/2d$, $T=0$, $B=0$. 
 }
\end{figure}

The positive background charge density and the sample width define the
characteristic energy $E_0$, Eq.~(\ref{vbg-inplane}). Measuring energies in
units of $E_0$, lengths in units of $d$, and density of states in units of
$D_0$, we obtain from Eq.~(\ref{thomas-fermi}) the dimensionless electron
density $\widetilde{n}(x/d)=n_{\rm el}(x)/E_0D_0$, so that Eq.~(\ref{hartree})
assumes a dimensionless form with the prefactor $1/\alpha_{\rm conf}$, where
\be \label{alpfa_conf}
\alpha_{\rm conf}=\pi a_B^{\star}/2 d
\ee
measures the relative strength of the Coulomb interaction, similar to $\alpha$
of Eq.~(\ref{alfascr}). We will usually assume $\alpha_{\rm conf}=0.01$, i.e.,
for GaAs, a sample width $2d\sim 3\, \mu$m, since this allows us to calculate
density profiles with clearly visible incompressible strips on a mesh of
relatively few ($\sim 500$) points across the sample. For much larger $d$, we
would need a much finer mesh, i.e., more ambitious numerics, and the
incompressible strips would be hardly visible on that scale, although the
physics would not change qualitatively.

Figure~\ref{fig:denspot} shows some density and potential profiles obtained
for the two sets of boundary conditions in the limit of zero temperature and
magnetic field, where $n_{\rm el}(x)/n_0=(\pi/\alpha_{\rm conf}) \mu(x)
\theta\big(\mu(x)\big)/E_0$, with $\mu(x)=\mu^{\star}-V(x)$. Apparently the
density profiles are very similar, if we  
assume the same depletion length, the same  sample width,  and vanishing
spacer between 2DEG and background charges (i.e., $z=0$) in both cases.

\subsection{Unmodulated system in a magnetic field}

\begin{figure}[h]
{\centering \includegraphics[width=\linewidth]{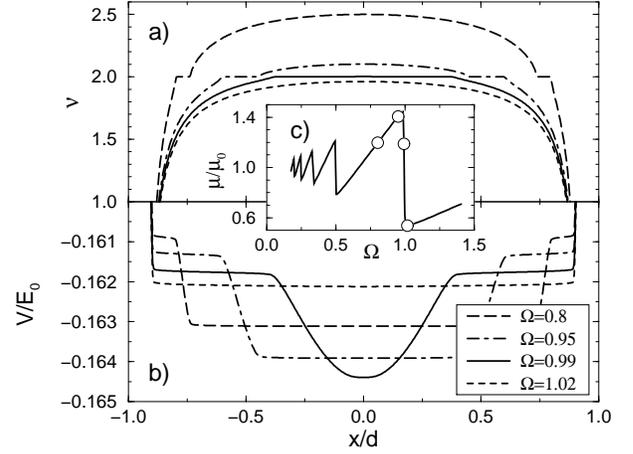}}
\caption{ \label{fig:unmod2}
a) Filling factor $\nu(x)$ and b) potential $V(x)$ for the values of $\Omega=
\hbar\omega_c/\mu_0$ given in the legend and indicated by open circles in c),
which shows the ``chemical potential'' $\mu=\mu^{\star}-V(0)$. Calculated with
model  (\ref{kernel-inplane}) for parameter values $\pi
a_B^{\star}/2d=0.01$, $\mu_0/E_0=0.00284$, $k_BT/E_0=2\times 10^{-5}$. 
 }
\end{figure}

In the ideal homogeneous 2DEG at low temperatures,
 the chemical potential as a function of
the magnetic field  exhibits  the well known saw-tooth behavior,
 Eq.~(\ref{sawtooth-mu}). With 
decreasing $B$ it follows a Landau energy $(n+1/2)\hbar \omega_c$ until the
 filling 
factor $\nu=2 E_F /\hbar \omega_c$ reaches the value $2(n+1)$, and then it
 jumps  to the next higher LL.
In the confined system, the self-consistently calculated ``chemical potential''
$\mu(0)=\mu^{\star}-V(0)$ in the center, $x=0$, shows the same behavior, as is
 seen in Fig.~\ref{fig:unmod2}c, where $\mu(x=0;B,T)$ in units of
 $\mu_0\equiv \mu(x=0;0,0)$ is plotted as function of $\Omega=\hbar
 \omega_c/\mu_0$.

 However, in contrast to the chemical potential oscillations
 in a homogeneous 2DEG, the corresponding oscillations in the confined system
 are realized by strong spatial variations of the electrostatic potential in
 the interior of the sample. This is demonstrated in  Fig.~\ref{fig:unmod2}b,
 which shows the self-consistent total potential in the interior of the sample
 for the four values of $\Omega$ indicated by open circles in
 Fig.~\ref{fig:unmod2}c. Figure~\ref{fig:unmod2}a shows the corresponding
 density profiles, normalized as local filling factor, $\nu(x)=2\pi l^2 n_{\rm
 el}(x)$. 
For $\nu(x) \leq \nu(0) \approx 2/\Omega <2$ the LL $n=0$ is pinned to the
 electrochemical potential $\mu^{\star}$ nearly everywhere in the 2DEG. For
 $\nu(0) \gtrsim 2$ the LL $n=1$ must be partially populated in the center of
 the sample. This forces the potential to develop a local minimum near the
 center, with a decrease  of $V(0)$ by an amount $\sim  \hbar\omega_c$, so
 that a compressible strip starts to develop in the center. With further
 increasing $\nu(0)$, this central compressible strip becomes broader and the
 adjacent incompressible strips, together with the related potential steps,
 move towards the sample edges. Similar drastic changes of the potential
 distribution are found near all jumps of the chemical potential. 
Thus, pinning and screening lead already to drastic effects in the confined
 2DEG  even in the absence of any additional potential modulation.

\subsection{Confined system with modulation}

We now add a symmetric external modulation potential to the confinement
potential and 
investigate how this affects the self-consistent potential. We take $V_{\rm
  ext}(x)=V_0\cos (2.5\pi x/d)$ which is in accord with our general boundary
conditions and exhibits just one full oscillation period in the interior of
the sample, so that we can expect similar screening effects as in a
homogeneous unbounded system, and possibly some effects of the nearby
sample edges. 
The period of this modulation is $a=2d/2.5$, so that the choice $\alpha_{\rm
  conf}=0.01$ implies $\alpha=1/40$ [see Eq.~(\ref{alfascr})] and the results
can be compared immediately with the previous one for the unbounded 2DEG.
For this comparison it will be important whether the potential of the
unmodulated system has a strong variation in the center region or not, i.e.,
whether the filling factor $\nu(0)$ 
in the center is slightly larger than an even integer or not.
\begin{figure}[h]
{\centering \includegraphics[width=\linewidth]{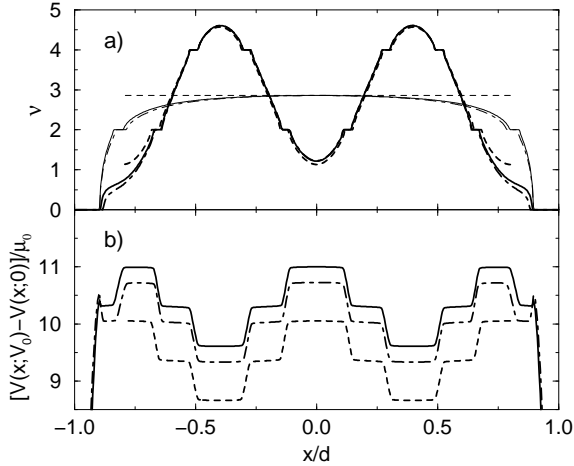}}
\caption{ \label{fig:comp3}
a) Filling factor $\nu(x)$ and b) screened potential $\Delta V(x;V_0)$  for
the confinement models  (\ref{kernel-inplane}) (solid lines) and
(\ref{perp-gates}) (dash-dotted lines), and for the unconfined 2DEG (dashed
lines). Thin lines show $\nu(x)$ without modulation, with
$\nu(0)=2.8574$ for all models. The screened potential for the unbounded 2DEG
is shifted by a constant, and actually oscillates symmetrically around zero.
($\alpha_{\rm conf}=0.01$, $\mu_0/E_0=0.002842$, $\hbar\omega_c/\mu_0=0.7$,
$V_0/\mu_0=24.63$, $k_BT/\mu_0=0.007$) }
\end{figure}
%
\subsubsection{Weak boundary effects on screening}
To describe the screening of an external modulation potential, it seems
natural to calculate the difference $\Delta V(x;V_0)=V(x;V_0)-V(x;0)$ of the
self-consistent potentials with and without the modulation. If $V(x;0)$ is flat
in the interior of the sample, we expect that screening is very similar to
that in an unconfined system and that $\Delta V(x;V_0)$ contains essentially
the same information as $V(x;V_0)$, apart from an unimportant constant offset.
This is indeed true if $\nu(0)$ is not closely above an even integer. As an
example, we compare in Fig.~\ref{fig:comp3}
numerical results for the two confinement models and for the unconfined 2DEG.
The results for the density of the confined 2DEG differ only slightly
in the edge regions. 
In the interior, the filling factors $\nu(x)$ are practically the same and
agree well with that of the unconfined 2DEG with the same modulation
potential. Also the screened potentials are equivalent and differ only by a
constant offset, which results from the asymmetry of the density modulation
with respect to the unmodulated electron density profile.

\subsubsection{Strong confinement effect on screening}

\begin{figure}[h]
{\centering \includegraphics[width=\linewidth]{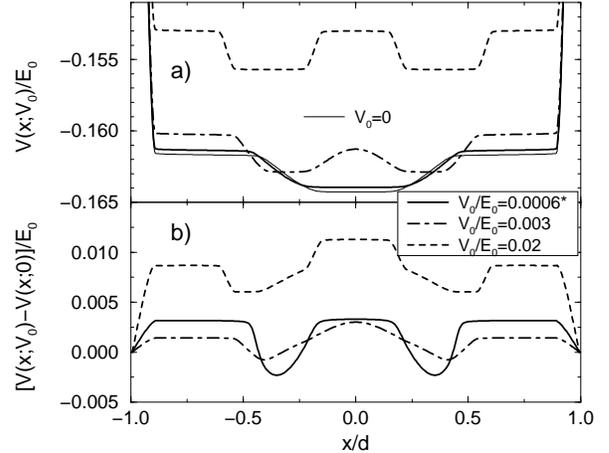}}
\caption{ \label{fig:var-vv03}
a) Selfconsistent potential $V(x;V_0)$ and b) screened potential $\Delta
V(x;V_0)$,  for several $V_0$. For clarity, b) shows $10\cdot \Delta V(x;V_0)$
for the weakest modulation $V_0=6\cdot 10^{-4} E_0$. For $V_0=0$ [thin solid
line in a)] $\nu(0)=2.03$.
($\alpha_{\rm conf}=0.01$, $\mu_0/E_0=0.002842$, $\hbar\omega_c/\mu_0=0.985$,
 $k_BT/\mu_0=0.01$) }
\end{figure}

Things become more complicated, if already without additional external
modulation the potential near the center of the Hall
bar   varies strongly, as happens for $\nu(0)=2n+\nu_n$ with $0<\nu_n \ll 1$.
Then, for small modulation amplitude ($V_0\ll \hbar \omega_c$), the
self-consistent potential 
$V(x;V_0)$ follows $V(x;0)$, with a minimum at $x=0$, and only the
difference $\Delta V(x;V_0)$ reminds of an oscillatory potential with the
phase of the external modulation,  see solid lines in Fig.~\ref{fig:var-vv03}.
For stronger modulation ($V_0 \lesssim \hbar \omega_c$),  $V(x;V_0)$ develops a
local maximum at $x=0$ and the total variation of $V(x;V_0)$ in the center
region $|x|\lesssim d/2$ is of the order of $V_0 < \hbar \omega_c$ see 
Fig.~\ref{fig:var-vv03}a. The variation of $\Delta V(x;V_0)$ is now, however,
by an amount of $\hbar \omega_c$ larger. In this small-$V_0$ regime screening
is rather poor and very nonlinear. 
As $V_0$ increases further, the variance ${\rm
  Var}[V](V_0)=V(0;V_0)-V(a/2;V_0)$ (note that $a/2=0.4\,d$) approaches the
  plateau value $\hbar\omega_c$,  and then behaves as a function of $V_0$ just
  as for the unconfined 2DEG. The variance of the ``screened potential'' 
$\Delta V(x;V_0)$, on the other hand, will be by about $ \hbar\omega_c$ larger
  in the plateau region. 

To summarize, if we neglect the relatively narrow $V_0$ interval between the
plateaus, we find that the variance of the self-consistent potential
$V(x;V_0)$ as a function of $V_0$ shows the same behavior as for the
unconfined 2DEG. For most magnetic field values, the variance of the 
``screened potential'' $\Delta V(x;V_0)$ also shows the same characteristics. 
Only if the filling factor $\nu(0)$ in the center of the unmodulated confined
2DEG is slightly larger than an even integer, the spatial variation of the
self-consistent potential $V(x;0)$ of the unmodulated system causes the
variance of  $\Delta V(x;V_0)$ to be about $\hbar \omega_c$ larger than the
variance of $V(x;V_0)$. If we plot the variance of  $\Delta V(x;V_0)$ at fixed
$V_0$ as a function of $\Omega=\hbar \omega_c/\mu_0$, we get sawtooth-like
traces as in Fig.~\ref{fig:varvsom}, however with additional spikes of height 
$\Omega$ at $\Omega\lesssim 1/k$ for integer $k$.

\section{Summary}

We have investigated the screening of a harmonic external potential by  an
unconfined two-dimensional electron gas as well as by confined 2DEGs in a
simplified Hall geometry, in strong perpendicular magnetic fields and at low
temperatures. 
Our numerical results within the self-consistent Thomas-Fermi-Poisson approach
show that screening is very nonlinear and dominated by the phenomenon of
pinning of Landau levels to the electrochemical potential, which leads to
compressible regions with position-dependent electron density, where this
pinning takes place, and to incompressible regions of constant density and
position-dependent electrostatic potential in between. At fixed magnetic
field, the total variation (``variance'' ${\rm Var}[V]$) of the
self-consistently calculated 
potential energy increases with increasing modulation amplitude $V_0$ in a
step-like fashion, exhibiting plateaus, where the value of ${\rm Var}[V]$ is
close to an integer multiple of the cyclotron energy and shows a weak linear
increase with  $V_0$, with a slope proportional to the temperature. The
corresponding modulation of the electron density is, in contrast to the
potential, not strongly affected by the magnetic $B$. The occurrence of
incompressible strips leads to local modifications, but the overall density
profile is roughly the same as for $B=0$, as has
already been emphasized by Chklovskii {\em et al.}\cite{Chklovskii92:4026}. 
Exploiting this observation together with the pinning phenomenon and the
relations between density modulation and external and screened potential
valid in the linear screening regime, we were able to derive simple analytical
expressions 
for step heights and plateau widths of the ${\rm Var}[V]$-vs-$V_0$
curves for  arbitrary $B$ and $T=0$. This simple analytical description of
nonlinear screening in an unconfined 2DEG is summarized in
Fig.~\ref{fig:slines}, and allows also an easy understanding of the
complicated traces obtained while plotting  ${\rm Var}[V]$ as a function of
$B$ at fixed $V_0$ (see Fig.~\ref{fig:varvsom}).

Finally we have investigated the corresponding screening properties of a
confined 2DEG in a simplified Hall geometry, for two different types of
boundary conditions, which lead to different confinement potentials, but
nearly identical density profiles, apart from slight deviations in the edge
regions. Considering the effect of an external modulation potential
$V_{\rm ext}(x)=V_0\cos (2.5 \pi x/a)$ in the interior of the sample (more
than about $a/2$ from the edges), we find essentially the same properties as
for the unbounded 2DEG. Care must however be taken if in the center of the
unmodulated system a new Landau level starts to be occupied, since then the
self-consistent potential varies strongly in the center region. This is an
interesting confinement effect, but
it can be easily eliminated from the discussion of screening if the modulation
amplitude $V_0$ is large enough.

\acknowledgments
We gratefully acknowledge  stimulating discussions with E.~Ahlswede and 
K.~Muraki as well as financial
support by the Deutsche Forschungsgemeinschaft, SP ``Quanten-Hall-Systeme''
GE306/4-2. 


\begin{thebibliography}{10}

\bibitem{Wulf88:4218}
U. Wulf, V. Gudmundsson, and R.~R. Gerhardts, Phys. Rev. B {\bf 38},  4218
  (1988).

\bibitem{Efros88:1019}
A.~L. Efros, Solid State Commun. {\bf 67},  1019  (1988).

\bibitem{Efros93:2233}
A.~L. Efros, F.~G. Pikus, and V.~G. Burnett, Phys. Rev. B {\bf 47},  2233
  (1993).

\bibitem{Burnett93:14365}
V.~G. Burnett, A.~L. Efros, and F.~G. Pikus, Phys. Rev. B {\bf 48},  14365
  (1993).

\bibitem{Cooper93:4530}
N.~R. Cooper and J.~T. Chalker, Phys. Rev. B {\bf 48},  4530  (1993).

\bibitem{Tsemekhman97:R10201}
V. Tsemekhman, K. Tsemekhman, C. Wexler, J.~H. Han, and D.~J. Thouless, Phys.
  Rev. B {\bf 55},  R10201  (1997).

\bibitem{Guven02:155316}
K. G{\"u}ven, R.~R. Gerhardts, I.~I. Kaya, B.~E. Sagol, and G. Nachtwei, Phys.
  Rev. B {\bf 65},  155316  (2002).

\bibitem{Kaya99:62}
I.~I. Kaya, G. Nachtwei, K. von Klitzing, and K. Eberl, Europhys. Lett. {\bf
  46},  62  (1999); and  Phys. Rev. B  (1998).

\bibitem{Kaya00:128}
I.~I. Kaya, G. Nachtwei, B.~E. Sagol, K. von Klitzing, and K. Eberl, Physica E
  {\bf 6},  128  (2000).


\bibitem{Chklovskii92:4026}
D.~B. Chklovskii, B.~I. Shklovskii, and L.~I. Glazman, Phys. Rev. B {\bf 46},
  4026  (1992).

\bibitem{Chklovskii93:12605}
D.~B. Chklovskii, K.~A. Matveev, and B.~I. Shklovskii, Phys. Rev. B {\bf 47},
  12605  (1993).

\bibitem{Lier94:7757}
K. Lier and R.~R. Gerhardts, Phys. Rev. B {\bf 50},  7757  (1994).

\bibitem{Oh97:13519}
J.~H. Oh and R.~R. Gerhardts, Phys. Rev. B {\bf 56},  13519  (1997).

\bibitem{Ando82:437}
T. Ando, A.~B. Fowler, and F. Stern, Rev. Mod. Phys. {\bf 54},  437  (1982).

\bibitem{Morse-Feshbach53:1240}
P.~M. Morse and H. Feshbach, {\em Methods of Theoretical Physics} (McGraw-Hill,
  New York, 1953), Vol.~II, p. 1240.

\bibitem{Stern67:546}
F. Stern, Phys. Rev. Lett. {\bf 18},  546  (1967).

\bibitem{Wulf88:162}
U. Wulf and R.~R. Gerhardts,  in {\em Physics and Technology of Submicron
  Structures}, Vol.~83 of {\em Springer Series in Solid-State Sciences}, edited
  by H. Heinrich, G. Bauer, and F. Kuchar (Springer-Verlag, Berlin, 1988), p.\
  162.

\bibitem{Gradshteyn}
I.~S. Gradshteyn and I.~M. Ryzhik, {\em Table of Integrals, Series, and
  Products} (Academic Press, New York, 1994).

\end{thebibliography}

\end{document}